\documentclass{cimento}
\usepackage{subcaption}

\usepackage{comment}

\usepackage{graphicx}  
\usepackage{float}
\usepackage{wrapfig}
\title{Status of the production of GEM chambers for the CMS experiment at Large Hadron
Collider}
\author{
L.~Benussi\from{ins:a},
S.~Bianco\from{ins:a},
R.~Campagnola\thanks{Corresponding author and presenter}\from{ins:a}\ETC,
M.~Caponero\from{ins:a}\from{ins:b},
S.~Colafranceschi\from{ins:c},
S.~Meola\from{ins:a}\from{ins:d},
E.~Paoletti\from{ins:a},
L.~Passamonti\from{ins:a},
D.~Piccolo\from{ins:a},
D.~Pierluigi\from{ins:a},
A.~Russo\from{ins:a},
G.~Saviano\from{ins:a}\from{ins:e},
R.~Tesauro\from{ins:a}
}
\instlist{\inst{ins:a} Istituto Nazionale di Fisica Nucleare, Laboratori Nazionali di Frascati, Frascati, Italy. ROR: 049jf1a25
  \inst{ins:b} ENEA Frascati, Italy. ROR: 01026pq66
  \inst{ins:c}Eastern Mennonite University, Harrisonburg, VA 22802, USA ROR: 059xmmg10
  \inst{ins:d}Università degli studi "G.Marconi", Rome, Italy. ROR: 00j0rk173
  \inst{ins:e}Sapienza Università di Roma, Rome, Italy. ROR: 02be6w209

  }

\begin{document}

\maketitle

\begin{abstract}

The High Luminosity LHC phase includes an upgrade to the muon stations for the CMS Experiment. CMS trigger and muon identification performance will be crucial, and it is, therefore, necessary to install new GEM stations to extend acceptance in the high-$ \eta $ region. An explanation of the quality control test and an update on the status of production will be provided. 
\end{abstract}



\section{Introduction} 

High Luminosity LHC (HL-LHC)  will increase the integrated luminosity up to 3000 $fb^{-1}$, tenfold compared to the design value of LHC, and instantaneous luminosity up to 5 - 7.5  $\cdot10^{34} {\rm cm^{-2}}{\rm s^{-1}} $  \cite{Apollinari:2284929}. To enhance CMS trigger and muon identification performance in high-rate environment, a second GEM (Gas electron multiplier)-based muon endcap station will be installed \cite{sauli1997gem}. CMS performance improvements aim to extend acceptance in the $1.6\; < \;\mid \eta \mid \;<\:2.4$ region\footnote{where $\eta$ is pseudorapidity, defined as: $\eta=-ln[tan(\frac{\theta}{2})]$, with $\theta$ the angle between the particle three-momentum $\mathbf {p}$ and the beam axis} by adding new detectors in GE2/1 position\cite{CERN-LHCC-2017-012} achieving a rate capability of $2\; {\rm kHz}/{\rm cm}^2$, a time resolution of $8-10\; {\rm ns}$ and an angular resolution of $500\;{\rm \mu} {\rm rad}$. 
The construction of GE2/1 station, begun in 2021, incorporates advanced quality controls and performance checks \cite{abbas2022quality}. 
 
 \begin{figure}[H]
\includegraphics[width=0.75\linewidth]{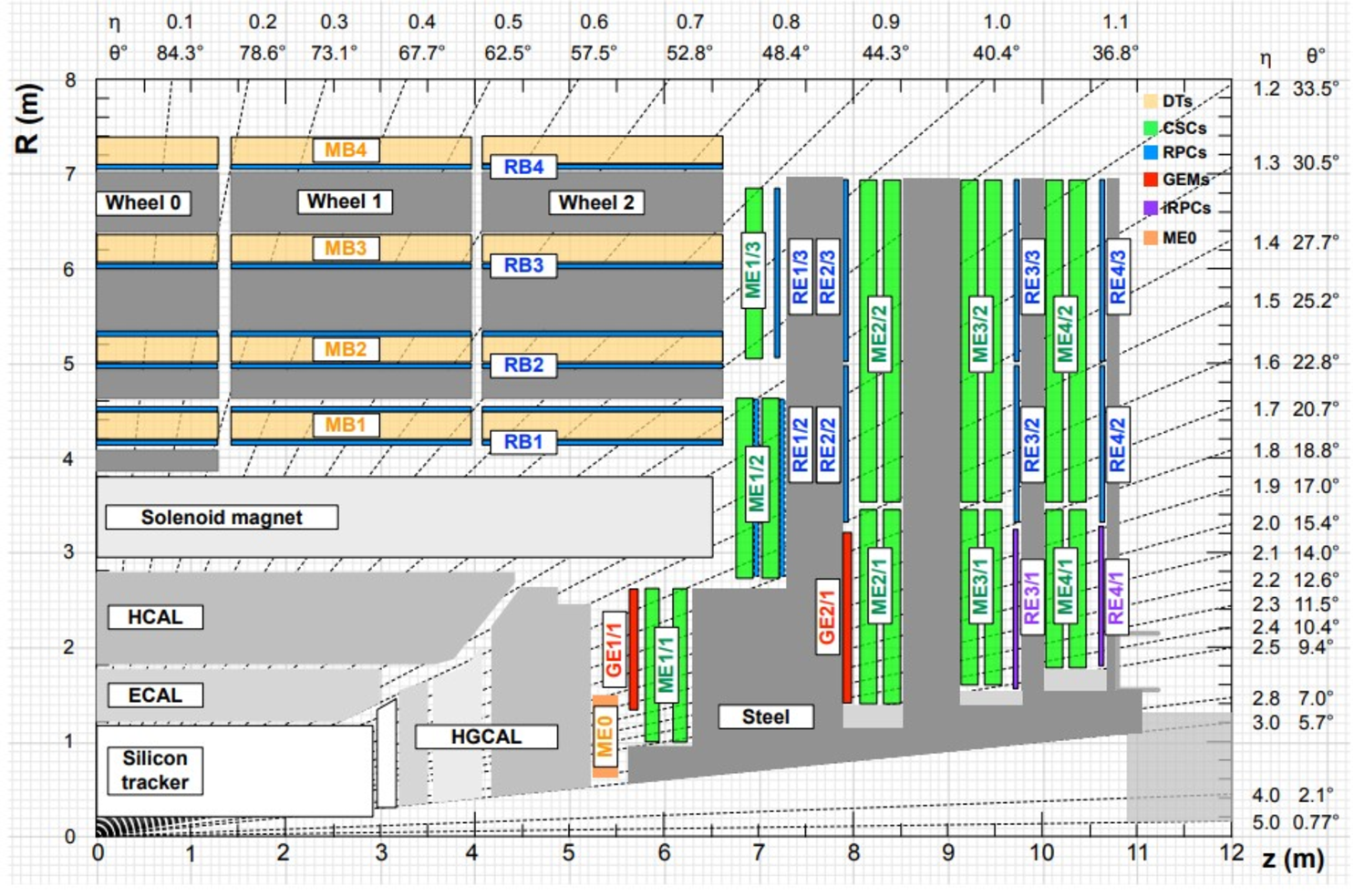}
\centering
\caption{Positioning of GE2/1 chambers in CMS}
\label{posizione}
\end{figure}
\section{Detector assembly and Quality Control}
A GEM is a $50\;{\rm \mu m} $ Kapton foil, metal-coated on both sides with a $5\;{\rm \mu m}$ copper layer and pierced with a high density of holes. With a voltage applied across the two sides of the foil, a strong electric field (60-100 kV/cm) is produced inside the GEM holes.  
By cascading the foils, it is possible to contain the amplification per individual stage, avoiding discharges while maintaining high gains (order of $10^{4}$).\\ 
The module for GE2/1 station consists of a stack of three GEM foils, enclosed between a 
drift and a readout electrode, held together by an external frame. The internal volume is filled with a mixture of ${\rm Ar}/{\rm CO_2}$ (70/30). 
Three GEM foils are stacked at their edges using thin internal frames. Foils are segmented in strips to protect them from irreversible damages, reducing the energy released by a possible discharge. Module ROBs (Readout Boards) are also divided in 12 sectors, each with 128 strips.

 Modules are assembled partly at CERN, partly at external production sites (Frascati, Bari, Aachen-Ghent, Beijing). The collaboration adopted this decentralized production approach to guarantee timely mass production with inherent redundancy.
All modules undergo rigorous Quality Control (QC) divided into 8 phases: QC1-2 at CERN, QC3-5 at CERN and external production sites, QC6-8 at CERN.

QC1 involves the preliminary inspection of the module main components: the foil surface conditions, the HV lines and the SMD resistors, the Printed Circuit Boards (PCBs) that host the drift and readout electrodes, the internal and external frames.\\
QC2 involves the qualification of the GEM foils, that are electrically cleaned from dust or chemical contaminants. The foils resistance is checked: if the final impedance is higher than $10 \,G\Omega$ the foil is accepted. Subsequently, the leakage current and discharge rate are monitored for several hours with the GEM foil stored stored in a $N_2$ box and powered with HV. The observed leakage current must be less than $20\, {\rm \mu} {\rm A}$ with HV between 100 and 600 V for 90 minutes, and less than $2\,{\rm \mu} {\rm A}$ with HV at 600 V for 14 hours \cite{kim2022production}.\\
After QC1 and QC2, construction kits are shipped to the production sites for assembly. 
After assembly, the first test to be performed is the QC3 gas tightness check. The pressure drop inside the module is monitored as a function of time. Efficient gas tightness also prevents contamination from external source that may affect the detector performance ($O_2$, other electronegative elements and dust can compromise charge amplification and foils integrity). The maximum allowable pressure drop is  $\Delta p(t) < 7 \, {\rm mbar}/{\rm h}$ \cite{abbas2022quality}.

QC4 aims to establish the I-V characteristic of the detector to identify any malfunctions and defects in the HV circuit that powers the module. 

The module is flushed with $CO_2$, powered up to 4900 V, and the current through the powering circuit is measured as a function of applied voltage. The test is considered successful if the I-V curve shows a linear behavior. Non-linearity would indicate a parasitic impedance, caused by a defect in the HV circuit or an issue with the GEM foils.

\begin{figure}[H]
\centering
  \begin{subfigure}{0.75\textwidth}
    \centering
    \includegraphics[width=0.82\linewidth]{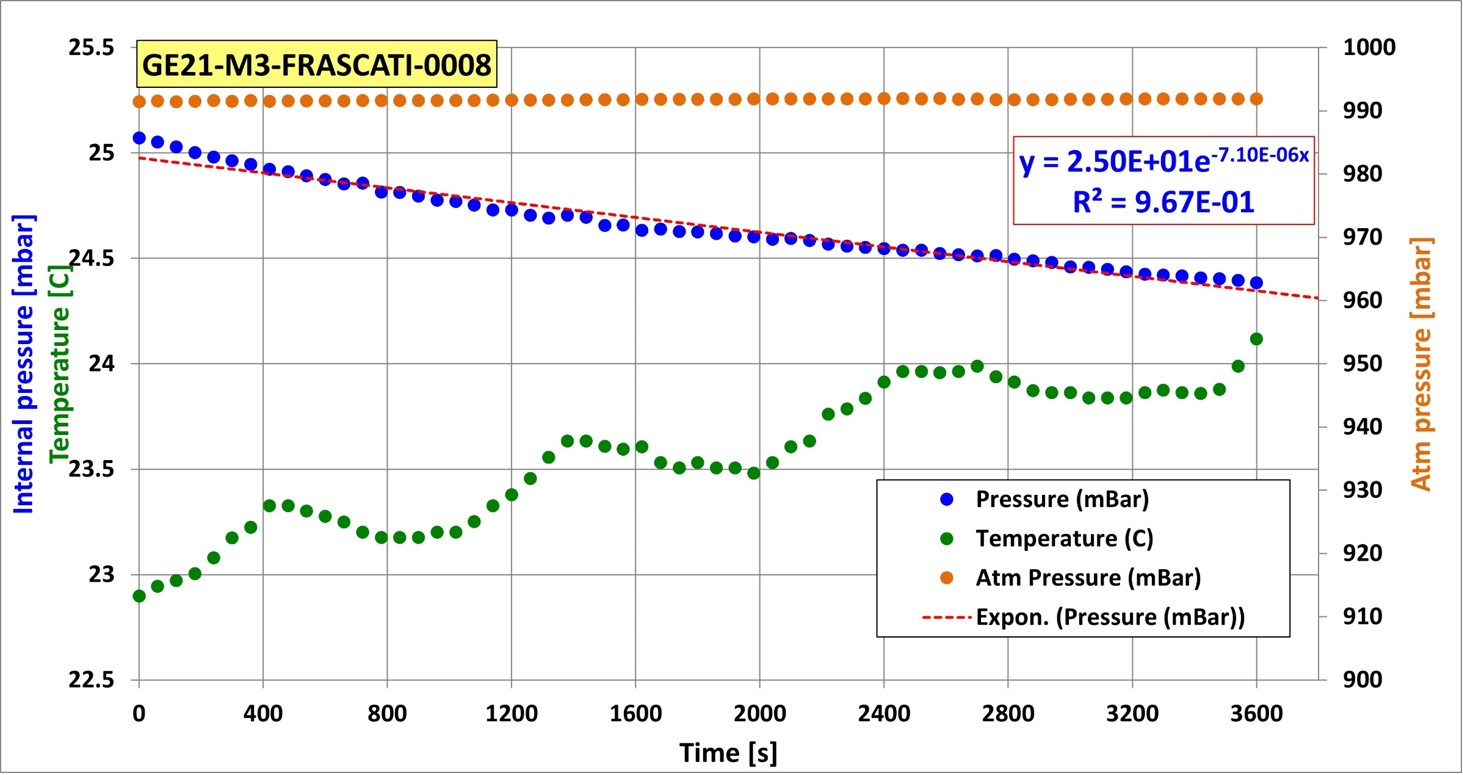}
    \caption{Internal over-pressure, atmospheric pressure and temperature  as a function of time.}
  \end{subfigure}
  \begin{subfigure}{0.75\textwidth}
    \centering
    \includegraphics[width=0.82\linewidth]{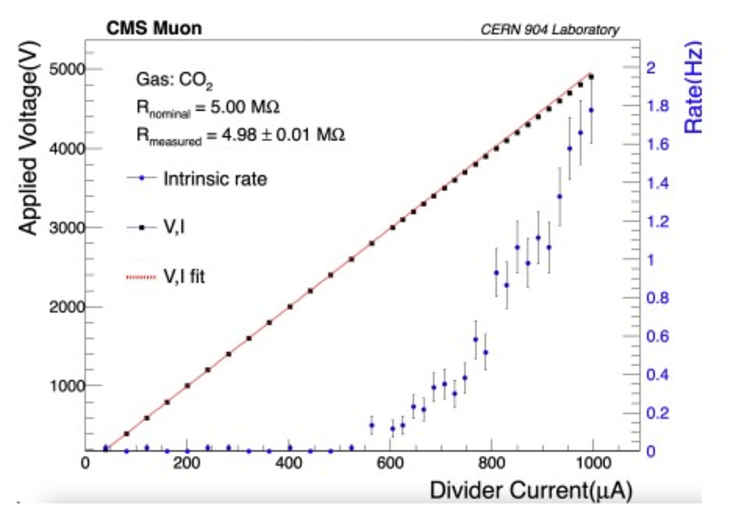}
    \caption{High voltage vs. divider current curve and spark rate for a GE2/1 detector \cite{kim2022production}}
  \end{subfigure}
  \caption{Results plot for QC3 and QC4.}
\end{figure}

The final QC5 step of in-site tests is divided into two parts and includes the effective gain measurement as a function of the divider current and gain uniformity test across the entire module.
An X-ray source is used to irradiate the whole module and generate a primary current, then subject to amplification; the test is conducted using an ${\rm Ar}/{\rm CO_2}$ mixture (70/30). The charge collected by each strip segmented on the ROB is measured. If the charge distribution has the same shape for each strip, then the detector is uniform.

\begin{figure}[H]
\includegraphics[width=0.79\linewidth]{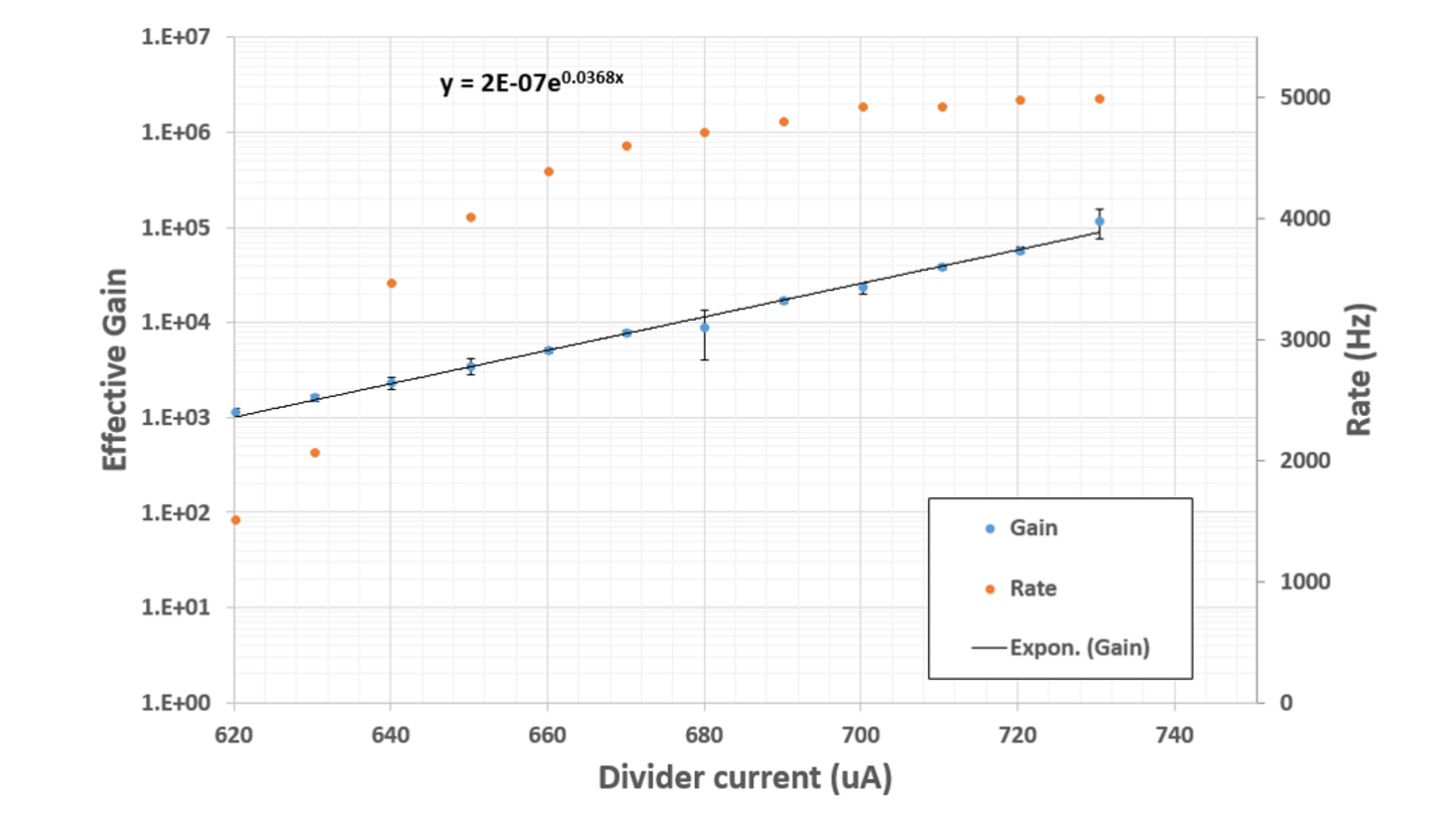}
\centering
\caption{Effective gain vs. divider current curve and spark rate for a GE2/1 detector}
\label{QC5}
\end{figure}
Once all test at the production sites are completed, the modules are shipped back to CERN for final quality tests on the electronics and validation of module performance (QC6, QC7, QC8).

\section{Results and conclusion}

A total of 288 modules (144 each endcap) is planned to be assembled. As this report is being prepared, 98 modules have been assembled (64 modules at CERN,
20 modules at Frascati and Bari sites). The CMS Frascati group has also contributed to the implementation of the FBG (Fiber Bragg Grating) sensors \cite{Benussi:2012owa} used for monitoring the operating temperature, a critical parameter for detector efficiency \cite{abbas2020two}. Production of the GEM modules will continue throughout 2024, followed by installation in the CMS detector.

\bibliographystyle{varenna}  
\bibliography{cimentobibliography.bib}

\begin{thebibliography}{1}
\expandafter\ifx\csname url\endcsname\relax\def\url#1{\texttt{#1}}\fi
\expandafter\ifx\csname urlprefix\endcsname\relax\def\urlprefix{URL }\fi

\bibitem{Apollinari:2284929}
\NAME{Apollinari G., Béjar~Alonso I., Brüning O., Fessia P., Lamont M., Rossi L. \atque Tavian L.}, \TITLE{{High-Luminosity Large Hadron Collider (HL-LHC): Technical Design Report V. 0.1}}, CERN Yellow Reports: Monographs (CERN, Geneva) 2017.
\newline\urlprefix\url{https://cds.cern.ch/record/2284929}

\bibitem{sauli1997gem}
\NAME{Sauli F.}, \IN{Nuclear Instruments and Methods in Physics Research Section A: Accelerators, Spectrometers, Detectors and Associated Equipment}{386}{1997}{531}.

\bibitem{CERN-LHCC-2017-012}
Tech. Rep. CERN, CERN, Geneva (2017).
\newline\urlprefix\url{https://cds.cern.ch/record/2283189}

\bibitem{abbas2022quality}
\NAME{Abbas M., Abbrescia M., Abdalla H., Abdelalim A., AbuZeid S., Agapitos A., Ahmad A., Ahmed A., Ahmed W., Aim{\`e} C. \etal}, \IN{Nuclear Instruments and Methods in Physics Research Section A: Accelerators, Spectrometers, Detectors and Associated Equipment}{1034}{2022}{166716}.

\bibitem{kim2022production}
\NAME{Kim M.~R.}, \TITLE{Production and quality control of the {GEM} {GE2/1} detector for the upgrade of the {CMS} endcap muon system,}, presented at \TITLE{PoS ICHEP2022 (2023)}, Vol. 414 2023.

\bibitem{Benussi:2012owa}
\NAME{Benussi L. \etal}, \IN{Phys. Procedia}{37}{2012}{483}.

\bibitem{abbas2020two}
\NAME{Abbas M., Abbrescia M., Abdalla H., Zeid S.~A., Agapitos A., Ahmad A., Ahmed A., Ahmed W., Amarjeet S., Asghar I. \etal}, \TITLE{Two years’ test of a temperature sensing system based on fibre bragg grating technology for the {CMS} {GE1/1} detectors}, in proc. of \TITLE{Journal of Physics: Conference Series}, Vol. 1561 (IOP Publishing) 2020, p. 012006.

\end{thebibliography}

\end{document}